\documentclass[aps,prl,twocolumn,floats,groupedaddress,showpacs, sort&compress, merge, superscriptaddress]{revtex4}%
\usepackage{amsmath}
\usepackage{amsfonts}
\usepackage{amssymb}
\usepackage{graphicx}
\usepackage{dcolumn}
\usepackage{natbib}

\newcommand{\la}{\langle}

\newcommand{\ra}{\rangle}

\begin{document}
\title{Parity Measurement is Sufficient for\\ Phase Estimation at the Quantum Cramer-Rao Bound for Path-Symmetric States}

\bibliographystyle{apsrev4-1}

\author{Sejong Kim}
\affiliation{Department of Mathematics, Louisiana State University, Baton Rouge, LA 70803}
\author{Kaushik P. Seshadreesan}
\email{ksesha1@lsu.edu}
\affiliation{Hearne Institute for Theoretical Physics and Department of Physics and Astronomy, Louisiana State University, Baton Rouge, LA 70803}
\author{Jonathan P. Dowling}
\affiliation{Hearne Institute for Theoretical Physics and Department of Physics and Astronomy, Louisiana State University, Baton Rouge, LA 70803}
\affiliation{Beijing Computational Science Research Center, Beijing, 100084, China}
\author{Hwang Lee}
\affiliation{Hearne Institute for Theoretical Physics and Department of Physics and Astronomy, Louisiana State University, Baton Rouge, LA 70803}
\date{\today}

\begin{abstract}
In this letter, we show that for all the so-called {\it path-symmetric} states, the measurement of parity of photon number at the output of an optical interferometer achieves maximal phase sensitivity at the quantum Cramer-Rao bound. Such optimal phase sensitivity with parity is attained at a suitable bias phase, which can be determined {\it a priori}. Our scheme is applicable for {\it local} phase estimation.
\end{abstract}

\maketitle


Interferometry is a vital component of various precision measurement, sensing, and imaging techniques. It works based on mapping the quantity of interest on to the unknown phase of a system and estimating the latter; for example, the relative phase between the two modes or ``arms" of an optical interferometer. Optical interferometry, often described in the Mach-Zehnder configuration, in general differs in the strategies of state preparation and detection. The conventional choice is to use a coherent light source and intensity difference detection. Assuming the unitary phase acquisition operator to be:
\begin{equation}
\label{evolution}
\hat{U}_\phi=e^{-i \phi(\hat{n}_a-\hat{n}_b)/2},
\end{equation}
(where $\hat{n}_a$, $\hat{n}_b$ are the number operators associated with the modes,) the phase sensitivity of the conventional Mach-Zehnder interferometer (MZI) is bounded by the shot noise limit (SNL) $\delta\phi=1/\sqrt{\bar{n}}$ (for $\bar{n}$ photons in the coherent state on average); whereas, protocols of quantum interferometry promise enhanced phase sensitivities by prescribing the use of states with nonclassical photon correlations and detection strategies that probe the particle nature of the output light, for example, via number counting~\cite{glm, *jonmetrology}. As for such nonclassical states, the proposal to squeeze the vacuum state entering the unused port of the conventional MZI, by Caves in 1981, resulted in the first instance of a quantum-enhanced MZI capable of operating below the SNL~\cite{caves}. Others, such as the twin-Fock state~\cite{holbur}, the maximally path-entangled N00N state $(|N,0\rangle+|0,N\rangle)/\sqrt{2}$, which reach the Heisenberg limit (HL) $\delta\phi=1/N$, were later proposed~\cite{barry, *bou, *jonlithography}.

Much of the latest experimental efforts in quantum interferometry have been focussed on attaining the HL~\cite{silberberg, *nagata, *higgins}. The theory of quantum phase estimation aids in identifying potential schemes for such phase sensitivities~\cite{sandersmilburn, *berrywiseman, *pezzi_2}. It is based on the information-theoretical concept of the Cramer-Rao bound~\cite{QCRB}. For the form of unitary phase acquisition considered in Eq.~(\ref{evolution}), the quantum Cramer-Rao bound (QCRB)---an attribute of the quantum state input to the MZI alone (independent of detection)---at its best, can reach Heisenberg scaling in the absence of photon losses. On the other hand, the classical Cramer-Rao bound (CCRB)---an attribute of the combination of a quantum state and detection strategy---can optimally reach the QCRB of the state. Uys and Meystre derived optimal quantum states, whose CCRB with number counting-based detection attains the HL in the absence of photon losses~\cite{Uys}. Lee {\it et al.} included photon losses to the problem and worked out the optimal inputs~\cite{taewoolee}. Meanwhile, Dorner {\it et al.} found optimal inputs in the presence of photon losses for the generic optimal detection strategy based on the symmetric logarithmic derivative operation, which is known to achieve the QCRB of all quantum states~\cite{dorner}. 
Pezz\'i and Smerzi revived Caves' original scheme, and showed that the QCRB of coherent mixed with squeezed-vacuum state input reaches the HL, $\delta\phi=1/\bar{n}$, when the two states are mixed in equal proportions ($\bar{n}/2$ photons on average in each state) at the input of the MZI~\cite{pezzismerzi2}. Additionally, they showed that the limit could be achieved with the scheme independent of the actual value of the phase, by gathering the full statistics of number counting in both the modes in place of intensity difference detection.

Apart from number counting and the symmetric logarithmic derivative operation, another detection strategy that has attracted a great deal of attention is the one based on the measurement of parity of photon number in one of the output modes of the MZI, described by the operator $\hat{\Pi}=(-1)^{\hat{n}}$~\cite{gerry2}. Parity detection was first proposed by Bollinger {\it et al.} for enhanced frequency measurement with an entangled state of trapped ions~\cite{bollinger, *Leibfried}. Later, Gerry and Campos applied parity measurement to optical interferometry with the N00N state for achieving phase sensitivities at the HL~\cite{gerry, *nonlinear}. Parity detection achieves sub-shot noise phase sensitivities with various inputs~\cite{aravindlee2}. MZIs with two-mode squeezed-vacuum state input and coherent state mixed with squeezed-vacuum state input have been shown to achieve the QCRB of the respective inputs with parity detection, which in turn can reach the Heisenberg limit~\cite{petr, kaushik1}. A theoretical question of interest is whether parity detection reaches the QCRB of all two-mode input states. In this letter, we prove that parity detection achieves the QCRB of all pure states that are {\it path-symmetric} inside the MZI~\cite{pathsymm}. The class of path-symmetric states includes all pure states that have a certain kind of coherence, i.e., phase relation---between the probability amplitudes in the photon number basis, as given in Eq.~(\ref{zbasiscond1}). All single mode state inputs to the MZI result in path-symmetric states after passing through the first 50:50 beam splitter. The N00N state and the states that result inside the MZI from inputs such as the twin-Fock state~\cite{holbur}, the Yuen state~\cite{yuen}, the two mode squeezed-vacuum state~\cite{petr}, the coherent state mixed with squeezed-vacuum state~\cite{hofmannono, *onoscheme, pezzismerzi2, kaushik1}, the pair-coherent state~\cite{pairgerry}, which are capable of HL phase sensitivity, are all path-symmetric states. Based on what we prove, it suffices to measure the parity of photon number in one of the output modes alone, in place of number counting in both the modes, in order to achieve the QCRB of all such path-symmetric states, some of which can in turn reach the HL.

\begin{figure}[h]\centering
\includegraphics[scale=0.40]{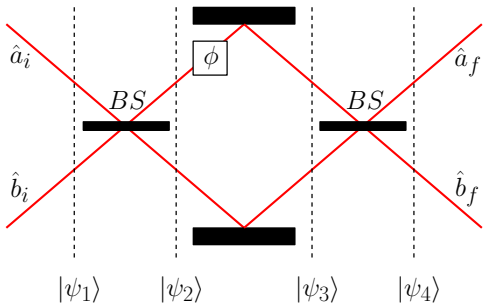}
\caption{A MZI with a two-mode input $| \psi_1\ra$, which after the beam splitter and phase shifter transformations $\hat{U}_{BS}=\textrm{exp}(-i\frac{\pi}{2}J_x)$, $\hat{U}_\phi=\textrm{exp}(-i\phi J_z)$ and $\hat{U}_{BS}^{\dagger}=\textrm{exp}(i\frac{\pi}{2}J_x)$ (in that order), is denoted by $| \psi_2 \ra$,  $|\psi_3 \ra$ and $|\psi_4 \ra$, at the respective stages.}
\label{fig:CohSV}
\end{figure}

We work in the Schwinger representation, wherein two-mode interferometry is described in terms of SU(2) algebra~\cite{yurke}. A two mode $N$-photon state in this representation resides in the $j=N/2$ subspace of the angular momentum Hilbert space. A typical MZI, $\hat{U}_{MZI}=\hat{U}_{BS}^{\dagger}\hat{U}_\phi\hat{U}_{BS}$, as described in Fig.~\ref{fig:CohSV}, is considered. For the chosen phase shifter and beam splitter transformations $\hat{U}_\phi=\textrm{exp}(-i\phi J_z)$ and $\hat{U}_{BS}=\textrm{exp}(-i\frac{\pi}{2}J_x)$, $\hat{U}_{MZI}=\textrm{exp}(-i\phi J_y)$. We call this the type-I MZI. On the other hand, if we choose a beam splitter that performs $\hat{U}_{BS}=\textrm{exp}(-i\frac{\pi}{2}J_y)$, then $\hat{U}_{MZI}=\textrm{exp}(-i\phi J_x)$, which we will call a type-II MZI. Without loss of generality, one can switch from type-I to type-II MZI by suitably adding a finite number of phase shifters before and after the beam splitters.

The error in the estimate of the unknown phase shift $\phi$ based on the measurement of an observable $\hat{O}$ can be written based on the linear error propagation formula as $\delta\phi=\left| \Delta O/(\partial \la \hat{O} \ra / \partial \phi) \right|$. 
For an observable that acts on the state $|\psi_3\rangle$, this quantity $\delta\phi$, due to the Heisenberg uncertainty principle between the generator of phase evolution $\hat{J_z}$ and the observable, obeys $\displaystyle (\delta \phi)  \geq
1/(2 \Delta J_z)$. 
For pure quantum states $|\psi_3\rangle$, the right-hand side of this inequality is a tight bound, and is identically equal to their QCRB. The equivalent (necessary and sufficient) condition for achieving the bound is that of equality in the Heisenberg uncertainty relation, namely:
\begin{equation} \label{Eq:equal-cond}
\displaystyle (\hat{O} - \la \hat{O} \ra I) | \psi_3 \ra = i \lambda (\hat{J_{z}} - \la\hat{J_{z}}
\ra I) | \psi_3 \ra,
\end{equation}
where $\lambda$ is purely real and the expectation values are calculated with respect to the state $|\psi_3\rangle$.

We first examine photon number counting-based detection strategies. For mathematical convenience, we choose a type-II MZI. Since photon counting at the output of a MZI is equivalent to the measurement of the operator $\hat{J_{z}}$ on the state $|\psi_4\rangle$, the corresponding observable $\hat{O}$ acting on $|\psi_3\rangle$ for such detection strategies must be diagonal in the $\hat{J_{x}}$ basis. This is so, because the $\hat{J_{z}}$ basis states transform into the $\hat{J_{x}}$ basis states under the action of the type-II beam splitter transformation $\hat{U}_{BS}=\textrm{exp}(-i\frac{\pi}{2}J_y)$. In the $\hat{J_{x}}$ basis $\{|m_x\rangle\}$, since the matrix elements of $\hat{J_z}$ are purely imaginary, the right hand side of Eq.~(\ref{Eq:equal-cond}) is real (Hermiticity requirement for the observable $\hat{O}$) if and only if the coefficients of $|\psi_3 \ra$ are purely real (up to a global phase factor), i.e.:
\begin{equation}
\label{xbasiscond}
\displaystyle \la m_x | \psi_3 \ra = \la m_x | \psi_3 \ra^{*} e^{-i 2 \chi},\ \forall\ m_x \in \{-j,...,+j\},
\end{equation}
where $\chi$ is a real constant, and $\la\hat{J_{z}}\ra=0$. This condition, when written in the $\hat{J_z}$ basis $\{|m_z\rangle\}$, transforms into:
\begin{equation}
\label{zbasiscond1}
\displaystyle \la m_z| \psi_3 \ra = \la -m_z | \psi_3 \ra^{*} e^{-i 2\chi},\ \forall\ m_z \in \{-j,...,+j\}.
\end{equation}
(Note that $\la\hat{J_z}\ra$ of such states is always zero.) Therefore, all pure states that obey Eq.~(\ref{zbasiscond1}) reach their maximal phase sensitivities with detection strategies based on photon number-counting. (For the rest of the paper, we will use $m\equiv m_z$.) 

Further, if $|\psi_2\rangle=\sum_{m=-j}^{+j}c_m|m\rangle$, then $|\psi_3\rangle=\sum_{m=-j}^{+j}c_m e^{-i m\phi}|m\rangle$, which implies:
\begin{equation}
\frac{\la m | \psi_3 \ra}{\la -m | \psi_3 \ra^*} =
\frac{c_m e^{-i \phi m}}{(c_{-m} e^{i \phi m})^{*}}
=\frac{ c_m}{c_{-m}^{*}}=\frac{\la m | \psi_2 \ra}{\la -m | \psi_2 \ra^*} .
\end{equation}
However, $\la m | \psi_3 \ra/\la -m | \psi_3 \ra^*= e^{-i 2\chi}$. Therefore $\la m | \psi_2 \ra/\la -m | \psi_2 \ra^*$ also equals $e^{-i 2\chi}$, i.e., the condition of Eq.~(\ref{zbasiscond1}) can be satisfied independently of the value of the unknown phase $\phi$. In a recent paper~\cite{pathsymm}, Hofmann interpreted this phase independence of Eq.~(\ref{zbasiscond1}) as a symmetry in the Heisenberg picture, wherein, the angular momentum vector ${\bf J}=\{J_x,J_y,J_z\}$ associated with the state is invariant under rotations about the $z$-axis~\cite{zrotation}. He called such states as {\it path-symmetric states}, since in the Schrodinger picture, the condition in Eq.~(\ref{zbasiscond1}) can be interpreted as the invariance of the two-mode pure state $|\psi\rangle$ with respect to a mode exchange operation (along with complex conjugation of the coefficients).

We now study such path-symmetric states in the context of parity-based metrology. Parity has been commonly studied with MZIs containing the beam splitter transformation $\hat{U}_{BS}=\textrm{exp}(-i\frac{\pi}{2}J_x)$; therefore we switch to the type-I MZI. The observable $\hat{O}$ (at the state $|\psi_3\rangle$) corresponding to parity measurement in the output mode $\hat{b}_f$ can be obtained by transforming $\hat{\Pi}=(-1)^{j-\hat{J_z}}$ through the beam splitter. It is given by the operator $\hat{Q}=\hat{U}_{BS}\hat{\Pi}\hat{U}_{BS}^{\dagger}$~\cite{aravindlee2}:
\begin{eqnarray}
\hat{Q} &=& i^{N} \sum_{k=0}^{N} (-1)^{k} | k,N-k \ra \la N-k,k | \\
&=& i^{N} \sum_{m = -j}^{j} (-1)^{j+m} | m \ra \la -m |,
\end{eqnarray}
and $\displaystyle \la\psi_4| \hat{\Pi} | \psi_4 \ra = \la\psi_3| \hat{Q} | \psi_3 \ra$. In order to achieve the quantum Cramer-Rao bound with the operator $\hat{Q}$ for a path-symmetric state $| \psi_3 \ra$, we need:
\begin{equation}
(\hat{Q} - \la \hat{Q} \ra I) | \psi_3 \ra = i \lambda \hat{J_z} | \psi_3 \ra,
\end{equation}
where $\displaystyle \la \hat{Q} \ra = \sum_{m=-j}^{j} i^{N} (-1)^{j+m} c_{-m} c_{m}^{*} e^{i 2m \phi}$  and $\displaystyle \lambda = \pm [\Delta{Q} / \Delta{J_z}]$ is a real number. Since $\hat{Q}^2=\hat{I}$, alternatively, we have
\begin{equation}
\frac{\la m | \hat{Q} | \psi_3 \ra - \la \hat{Q} \ra \la m | \psi_3 \ra}{\la m | \hat{J_z} | \psi_3 \ra} = \pm i \left[ \frac{1 - \la \hat{Q} \ra^2}{\la \hat{J_z}^{2} \ra} \right]^{1/2}.
\end{equation}
Via direct computation of each term, we obtain:
\begin{equation} \label{Eq:Equi-cond}
S- \la \hat{Q} \ra = \pm i m \left[ \frac{1 - \la \hat{Q} \ra^2}{\la \hat{J_z}^{2} \ra} \right]^{1/2},
\end{equation}
where $S=i^{N} (-1)^{j+m} \frac{c_{-m}}{c_{m}} e^{i 2m \phi}\equiv S'+i S''$. Let $c_{m} = r_{m} e^{i \theta_{m}}$, where $r_m$ and $\theta_m$ are real. Then,
\begin{equation} \label{Eq:cond-R}
\begin{split}
S'& = \textrm{Re} \left( i^{N} (-1)^{j+m} e^{i 2(m \phi - \chi - \theta_{m})} \right) \\
& = \begin{cases}
    \pm (-1)^{j+m} \sin2(m \phi - \chi - \theta_{m}) & \hbox{if $N$ is odd,} \\
    (-1)^{m} \cos2(m \phi - \chi - \theta_{m}) & \hbox{if $N$ is even.} \\
\end{cases}
\end{split}
\end{equation}
If $S' = \pm 1$ for some $\phi$ and for all $m$, then $S''=0$ since $\displaystyle S'^2 + S''^2 = \left| i^{N} (-1)^{j+m} e^{i 2(m \phi - \chi - \theta_{m})} \right|^2 = 1$. In that case, $\displaystyle \textrm{Re} \left( i^{N} (-1)^{j+m} c_{-m} c_{m}^{*} e^{i 2m \phi} \right) = \pm | c_{m} |^{2}$ and
$\displaystyle \textrm{Im} \left( i^{N} (-1)^{j+m} c_{-m} c_{m}^{*} e^{i 2m \phi} \right) = 0$ for all $m$. So
\begin{displaymath}
\displaystyle \la \hat{Q} \ra = \sum_{m=-j}^{j} i^{N} (-1)^{j+m} c_{-m} c_{m}^{*} e^{i 2m \phi} = \pm \sum_{m=-j}^{j} | c_{m} |^{2} = \pm 1,
\end{displaymath}
which implies that the condition in Eq.~(\ref{Eq:Equi-cond}) is satisfied, and hence the quantum Cramer-Rao bound is achieved. Therefore, $S'=\pm 1$ is a sufficient condition for a path-symmetric state to achieve its quantum Cramer-Rao bound with parity measurement.

Let us assume that phase sensitivity at the quantum Cramer-Rao bound is obtained for a path-symmetric state $| \psi_2 \ra$; i.e., the equation (\ref{Eq:Equi-cond}) holds for all $m$;
\begin{equation}
\displaystyle S' + i S''   - \la \hat{Q} \ra = \pm i m \left[ \frac{1 - \la \hat{Q} \ra^2}{\la \hat{J_z}^{2} \ra} \right]^{1/2}.
\end{equation}
Then, we have
\begin{equation}
\displaystyle S'= \la \hat{Q} \ra, \ S''= \pm m \left[ \frac{1 - \la \hat{Q} \ra^2}{\la \hat{J_z}^{2} \ra} \right]^{1/2},
\end{equation}
and furthermore,
\begin{equation}
\displaystyle S'' = \pm m \left[ \frac{1 - S'^2}{\la \hat{J_z}^{2} \ra} \right]^{1/2} = \pm m S'' \left[ \frac{1}{\la \hat{J_z}^{2} \ra} \right]^{1/2}.
\end{equation}
If $\left| S' \right| \neq 1$, then $S'' \neq 0$ since $S'^2 + S''^2 = 1$. So we get $\la \hat{J_z}^{2} \ra^{1/2} = \pm m$ for all $m$, but this result is a contradiction in general, since it should hold for all $m$. Hence, $\left| S' \right| = 1$.  Therefore, $S' = \pm 1$ is also a necessary condition for a path-symmetric state to achieve its quantum Cramer-Rao bound with parity measurement. Hence, we say {\it an $N$-photon path-symmetric state $| \psi_2 \ra= \sum_{m=-j}^{j} c_{m} | m \ra$ reaches its quantum Cramer-Rao bound with the parity detection operator $\hat{Q}$ if, and only if, there exists a real number $\phi$ such that, for all $m$,
\begin{equation}
\label{claim1}
\displaystyle {\rm Re} \left( i^{N} (-1)^{j+m} \frac{c_{-m}}{c_{m}} e^{i 2m \phi} \right) = \pm 1.
\end{equation}
}

Although seemingly complicated at the outset, the condition in Eq.~(\ref{claim1}) is easily satisfied in the following cases:
\begin{itemize}
\item[(a)] $c_m=r_m$, i.e., $c_{m}$ is purely real ($\theta_{m} = 0$) for all $m$, and therefore $\chi = 0$.
\item[(b)] $c_m=r_m e^{i m \theta}$, i.e., $\theta_m=m \theta$ for some constant $\theta$, and therefore $\chi = 0$.
\end{itemize}
The values of $\phi$ in case (a), and $\phi - \theta$ in case (b), when chosen as:
\begin{equation} \label{Eq:cond-phi}
\displaystyle \begin{cases}
    \frac{\pi}{2} \ \textrm{or} \ \frac{3 \pi}{2} & \hbox{if $N$ is odd,} \\
    0 \ \textrm{or} \ \frac{\pi}{2} \ \textrm{or} \ \pi \ \textrm{or} \ \frac{3 \pi}{2} & \hbox{if $N$ is even,} \\
\end{cases}
\end{equation}
satisfy the condition in Eq.~(\ref{claim1}), and consequently result in maximally sensitive phase estimation with parity measurement for the given state at those values of $\phi$.

Additionally, given any arbitrary path-symmetric state $|\psi_2\ra$ with real parameters $\{r_m, \theta_m\}$, and a real constant $\chi$, there always exists a bias phase $\beta$, such that $\theta_m+m\beta=-\chi$ for all $m$; and the value of $\beta$ can be obtained from:
\begin{equation}
\label{biasphase1}
\sum_{m>0}^{j}(\theta_m+m\beta)= -\sum_{m>0}^{j}\chi.
\end{equation}
Solving Eq.~(\ref{biasphase1}), we get:
\begin{equation} \label{biasphase2}
\displaystyle \beta=\begin{cases}
   -\frac{8}{(N+1)^2}(\frac{(N+1)}{2}\chi+\sum_{m>0}^{j}\theta_m) & \hbox{if $N$ is odd,} \\
   -\frac{8}{N(N+2)}(\frac{N}{2}\chi+\sum_{m>0}^{j}\theta_m) & \hbox{if $N$ is even.} \\
\end{cases}
\end{equation}
By replacing $-\chi-\theta_m$ in Eq.~(\ref{Eq:cond-R}) with $m\beta$, and choosing the value of $\phi-\beta$ as per Eq.~(\ref{Eq:cond-phi}), the QCRB of the input state can be reached with the operator $\hat{Q}$ at the corresponding value of $\phi$. So, {\it for any given arbitrary $N$-photon path symmetric state, there always exists a real phase $\phi$, wherein parity detection attains the QCRB of the state.} These results directly extend to states with indefinite photon numbers, with the optimal point of operation (the value of $\phi$ or $\phi-\theta$ or $\phi-\beta$, as is the case) given by the values common to the even and odd components of Eq.~(\ref{Eq:cond-phi}), i.e. $(2l+1)\pi/2, \ l\in \mathbb{Z}$.

Although optimal phase sensitivities with parity are reached only at specific bias phases, yet, since the bias can be predetermined, parity detection provides a useful tool for ``local" phase estimation. The biases may be easily implemented in the laboratory if we notice that the phase $\phi$ is the relative phase between the two arms of the interferometer, and hence can be always tuned to the optimal bias point or ``sweet spot", by inserting a an adjustable and known phase shift in the other arm. For detecting small changes of more or less known parameters over space or time, {\it a priori} knowledge about the parameter is assumed---this is called local parameter estimation---as opposed to the ``global" one, wherein a complete lack of knowledge about the parameter is initially assumed~\cite{durkin}.

One way to measure parity is obviously to perform number counting at the output and to infer the parity from it. High precision number-resolving detection of photons at single-photon level has been demonstrated experimentally with the superconducting transition-edge sensors~\cite{pnr0}. Apart from number counting, there have been other methods proposed for the direct measurement of parity. Gerry and co-workers suggested the use of optical nonlinearities~\cite{gerry, *nonlinear}. Plick {\it et al.} showed that homodyne quantum state tomography can be used to construct the expected parity signal, at least in the case of Gaussian states, since the expectation value of the parity operator is proportional to the value of the Wigner function of the state at the origin in phase space for such states~\cite{royer, *bill}. 

In conclusion, we have shown that the detection strategy based on the measurement of the parity of photon number achieves the maximal phase sensitivity at the quantum Cramer-Rao bound of all pure states that are path-symmetric. Although such optimal phase sensitivities with parity detection are achieved only locally, those specific bias phases can be predetermined.

The authors wish to thank P. M. Anisimov for stimulating discussions. This work was supported by the National Science Foundation.

\bibliography{references_1}

\begin{thebibliography}{37}
\bibitem [{zro()}]{zrotation}%
  \BibitemOpen
  \href@noop {} {}\bibinfo {note} {This is because phase shifts are represented
  by $J_z$ rotations in the Schwinger representation.}\BibitemShut {Stop}%
\end{thebibliography}
\end{document}